"Quantum supremacy" revisited: Low-complexity, deterministic solutions of the *original* Deutsch-Jozsa problem in classical physical systems

**Laszlo Bela Kish**

*Department of Electrical and Computer Engineering, Texas A&M University, TAMUS 3128, College Station, TX 77843-3128, USA*

**Keywords:** The *original* Deutsch-Jozsa problem; Instantaneous noise-based logic; classical databases; computational complexity.

*Author for correspondence (Laszlokish@tamu.edu).
†Present address: Department of Electrical and Computer Engineering, Texas A&M University, TAMUS 3128, College Station, TX 77843-3128 USA

# 1. Summary

The *original* Deutsch-Jozsa (oDJ) problem is for an oracle (realized here as a database) of size *N*, where, according to their claim, the deterministic solution of the problem on a classical Turing computer requires O(*N*) computational complexity. They produced the famous Deutsch-Jozsa quantum algorithm that offered an exponential speedup over the classical computer, namely O[log(*N*)] complexity for the solution in a quantum computer. In this paper, the problem is implemented on an instantaneous noise-based logic processor. It is shown that, similarly to the quantum algorithm, the oDJ problem can deterministically be solved with O[log(*N*)] complexity. The implication is that by adding a truly random coin to a classical Turing machine and using this classical-physical algorithm can also speed up the deterministic solution of the Deutsch-Jozsa problem exponentially, similarly to the quantum algorithm. Then it is realized that the same database and the solution of the Deutsch-Jozsa problem can also be realized by using an identical algorithmic structure in a simpler way, even without noise/random coin. The only lost function in this new system, as compared to noise-based logic, is the ability to do generic parallel logic operations over the whole database. As the latter feature is not needed for the oDJ problem, it is concluded that the problem can be solved on a classical computer with O[log(*N*)] complexity even without a random coin. Therefore, while the oDJ algorithm is historically important step in the developments of quantum computers, it is insufficient to prove quantum supremacy. Note, there is also *simplified* Deutsch-Jozsa problem proposed later, which is more popular in the field, however it is irrelevant for the present paper.

# 2. Introduction

There is a growing discussion about the so-called "quantum supremacy" over classical Turing machines, see e.g. [1-4]. Quantum supremacy means that there exist specific, "hard" mathematical problems where the solutions on a classical-physical Turing machine require exponentially larger computational complexity (hardware and/or time) compared to a quantum computer. That typically means an exponential speedup at identical hardware complexity. The first breakthrough to show quantum supremacy is the oDJ quantum algorithm [5] that produces a deterministic solution that is exponentially faster than a classical algorithm on an envisioned classical processor. In this paper, first we show a deterministic algorithm running on an instantaneous noise-based logic (INBL) processor, which requires a classical-physical Turing machine that is expanded by a true random number generator (TRNG, truly random coin). Its performance regarding the oDJ problem is in the same complexity class: the classical algorithm is similarly fast as the quantum algorithm, while it is also deterministic. Then we simplify that system and show an alternative solution, with the same logic structure but without a TRNG, that has the same complexity class.

*2.1. The original Deutsch-Jozsa problem*

We formulate the original Deutsch-Jozsa problem [5] in the following way (by using *N* instead of 2*N* that was used in [5]). Suppose there is a database $\{f_i\}$ with *N* single-bit elements, $f_i \in \{0,1\}$, where (*i*= 1,2,...*N*) and, for convenience, $N = 2^M$, where M is the number of qubits or noise-bits. Find which of the following statements are true:

(2.a) The database is *not* a uniform set of (0 or 1) bit values.

(2.b) The database does *not* contain exactly *N*/2 bit values of zero.

It is important to note here that the oDJ problem statement above and in [5] allows *arbitrary* situations because DJ explicitly states there [5]:

(2.c) "*Note that for any f, at least one of (2.a) and (2.b) is always true. It may be that both are true, in which case either (A) and (B) is an acceptable solution*".

Subsequently, the DJ problem was modified to today's popular version, see for example [6], which we call *simplified* DJ problem; by adding the following condition:

(2.d) *"We are promised that the function is either constant (0 on all inputs or 1 on all inputs) or balanced (1 for exactly half of the input domain and 0 for the other half)"* [6].

Obviously, the simplified DJ problem is limited because (2.c) and (2.d) directly contradict each other. The solution in the present paper is for the oDJ problem [5] and it is irrelevant for its simplified version [6].

The oDJ quantum algorithm [5] can solve this problem on a *full* quantum computer (that is, *M* qubits with $N=2^M$ dimensional Hilbert space) with polynomial-in-*M*, that is, O[log(*N*)] complexity. The theoretical solution is by (typically unitary) operations over the quantum superposition representing the database. At the hardware side, the physical realization of such operations in a quantum computer include switching control elements (for example, polarizers; beam-splitters; switches between Josephson junctions; magnetic or electrical field, etc.); inputting parameters/data; and executing measurements. (This situation is similar also with noise-based logic, see below).

An important condition for the usefulness of a quantum computer for a given problem is that the required hardware manipulations for the solution should be limited to polynomial-in-*M* hardware and time complexity, even if the superposition carrying the data system itself represents data with exponential complexity. To illustrate this situation, we use an example where *Alice* (implementer) and *Bob* (solver) are the key players, see Figure 1.

(i) Alice implements a problem in the quantum computer. Depending on the input, she sets up the superposition with her hardware that provides the initial and boundary conditions for the wavefunction carrying the superposition representing the data. Note, Alice may need resources (hardware and/or time) with O($N=2^M$) complexity when she carries this procedure out. For example, filling a database with $N=2^M$ *arbitrary* bit values requires O(*N*) complexity, even though some special tasks (such as filling in special types of data sets, e.g. pure zeros, ones, or a superposition of all integer numbers - the *universe*, etc.) may need only polynomial-in-*M* complexity. Thus, when Alice prepares the database on which the oDJ algorithm [5] will run, she must use exponential-in-*M*, O(*N*), complexity.

(ii) Bob carries out the computation that, in our example, is the oDJ algorithm [5], to solve the problem and to output the results. He may execute the operations on a superposition by using some of the hardware components of Alice combined with his own extra hardware. As we mentioned above, the important feature of a useful quantum processor is that Bob's required computational complexity to obtain the results is only polynomial-in-$M$.

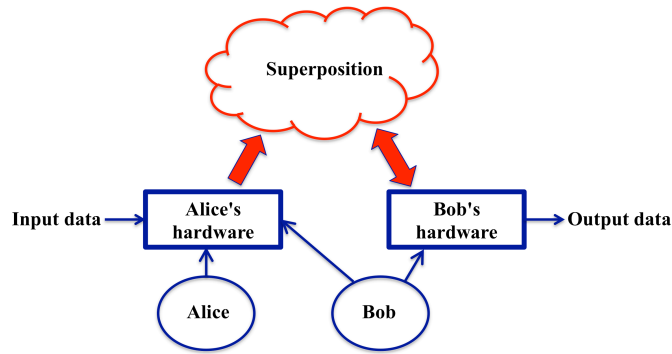

**Figure 1.** Outline of a generic quantum computer and its operation in a problem *implementer* (Alice) and *solver* (Bob) fashion. Alice implements the problem in the quantum computer and at certain tasks she may need a complexity that is exponential in $M$ (e.g. when filling in exponentially large, $N=2^M$, arbitrary databases). Depending on the input, she sets up the superposition with her hardware that provides the initial and boundary conditions for the wavefunction carrying the superposition that represents the data. Bob carries out the computation to solve the problem and does the measurements to output the results. He may execute the operations on the superposition by using some of the hardware components of Alice and his own extra hardware. The important feature of a useful quantum processor is that Bob's required computational complexity to obtain the results is only polynomial-in-$M$. The noise-based logic implementation of the problem has similar features, except that the superposition is classical, moreover nonlinear processes (products) and a true random number generator are also involved in the process.

The oDJ algorithm is not described here but it is emphasized that its complexity is $O(M) = O[\log(N)]$ while it serves a *deterministic* solution. According to [5], the algorithm represents "quantum supremacy" over purely classical physical Turing computers because in the envisioned classical computer and algorithm of [5], an exponentially larger, $O(N) = O[2^M]$, complexity is required for the deterministic solution. We will show in Sections 3 and 4 that there are low-complexity deterministic solutions of the oDJ problem in a classical physical system that challenge the "quantum supremacy" claim.

As a preparation to the classical solutions, instantaneous noise-based logic is introduced below.

*2.2 Instantaneous noise-based logic*

Noise-based logic (NBL) [7-22] is a classical physical computational scheme, where the logic information is carried by (often the superposition of) independent, stationary stochastic processes (noises). The bit values are represented by stationary, orthogonal (uncorrelated) noises with zero mean values. These noises can be viewed as the fingerprints of these bit values. A pair of noises representing the High (1) and Low (0) values of a bit is called *noise-bit* (similarly to the *qubit* of quantum computing). Naturally, for an NBL processor with $M$ noise-bits, $2M$ independent, stationary, truly random noises with zero mean values are required. Figure 2 shows the generic outline of NBL. All the $2M$ noise-bit values are generated by the Reference Noise System (RNS) and they are distributed as reference signals all over the NBL processor. Utilizing the received input logic information and the reference signals, the NBL gates determine their output signal that is also a noise that can be: any of the reference signals; their arbitrary superposition; their products; or the arbitrary superposition of their products. The nonlinear operations represented by the products of noises allow an exponentially large, $2^M$ dimensional Hilbert space [7,8] (see below). Logic operations can be executed by not only the gates but also by operations on the RNS.

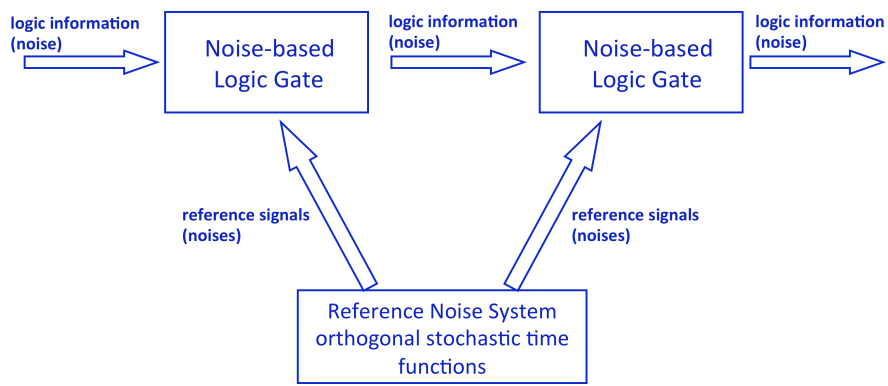

**Figure 2.** Generic noise-based logic hardware scheme [7-22]. Logic operations can be executed by the gates or by operations on the reference signals. The Reference Noise System is based on a truly random number generator. The interesting concrete schemes and algorithms are those can be implemented by a binary classical computer (Turing machine) with a bit resolution that is polynomial in $M$., e.g. [15-22].

There are many types of NBLs depending on the goals and applications, and here we are focusing on instantaneous noise-based logic (INBL, [9,10]), which has many sub-types regarding the kind of RNS and its internal timing [11-15]. Regarding our notations, a reference system of $M$ noise-bits with the $2M$ true random noise generators is given as:

$$R_{10}(t), R_{11}(t), R_{20}(t), R_{21}(t), ..., R_{M0}(t), R_{M1}(t) \qquad (1)$$

where the first index is the bit number and the second one represents the Low (0) and High (1) values of that noise-bit. The reference noises (in the infinite-time limit) can also be viewed as *orthogonal vectors*:

$$\langle R_{ij}(t)R_{pq}(t)\rangle = 0 \quad \text{for} \quad i \neq p \text{ and/or } j \neq q, \tag{2}$$

where $\langle \ \rangle$ means time average.

Thus, superpositions of these signals represent a $2M$ dimensional geometrical *space*.

However, product operations between the reference noises and their superpositions can lead out from this original space and form a *hyperspace* [7] which is also a Hilbert space, with exponentially high ($2^M$) dimensions [7]. In this paper, we use the scheme that was used for the "drawing from hats" [20] and the "NBL phonebook search" [21,22] systems, which is the so-called *instantaneous* NBL (INBL). See the circuitry illustration in Figure 3. The $2M$ different RNS signals are fed into the Hilbert Space Synthesizer, which is a binary classical processor of O($M$) (polynomial) bit resolution (the exact resolution requirement depends on the application). This system can conveniently be realized by a digital computer, chip or other circuitry, and an additional TRNG. The output signal represents the generated special-purpose superposition that can be used for further processing.

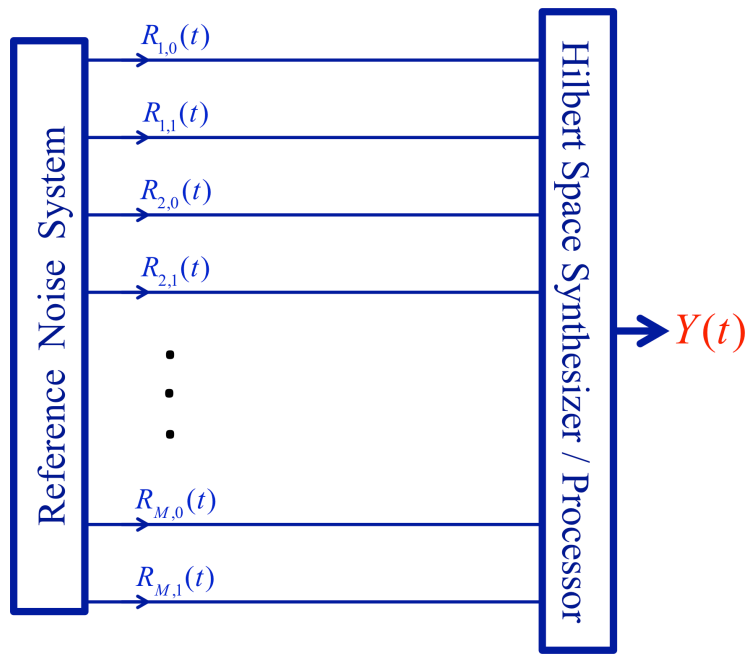

**Figure 3.** Circuit illustration of the logic structure of the generic superposition synthesizer for instantaneous NBL (INBL). For examples, see the text.

Some examples:

In INBL systems, binary numbers are $M$-bit long product-strings formed by the noises of the corresponding reference signals. For example, in a 3 noise-bit system, $M = 3$, the signal $W_6(t)$ of the number 6 and its binary version, 110, is carried by the noise product

$$W_6(t) = R_{10}(t)R_{21}(t)R_{31}(t), \tag{3}$$

where bit-1 is the bit with the lowest numeric value (0 or 1), and bit-3 is with the highest numeric value (0 or 4).

In a 4-noise-bit system, the signal $Y_{7,4,1}(t)$ of the superposition of the numbers 7 (that is, 111), 4 (that is, 100) and 1 (that is 001) is given as:

$$Y_{7,4,1}(t) = W_7(t) + W_4(t) + W_1(t) = R_{11}(t)R_{21}(t)R_{31}(t) + R_{10}(t)R_{20}(t)R_{31}(t) + R_{11}(t)R_{20}(t)R_{30}(t). \tag{4}$$

The product strings belonging to different numbers are orthogonal due to Equation (2), so their system represents a space with $2^M$ dimensions.

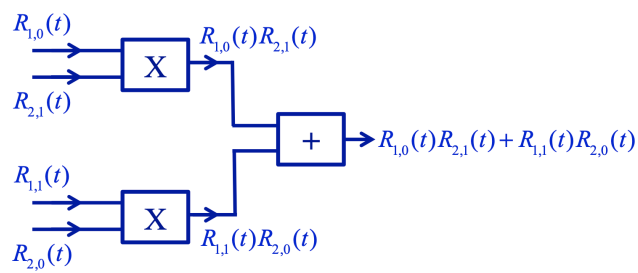

**Figure 4.** Simple circuit example utilizing multipliers and adders as an example for the Hilbert Space Synthesizer: The Bell state in a 2 noise-bit system [21]. Arbitrary exponential superpositions can require exponential hardware complexity. Thus, similarly to quantum computers, in general applications, usually those exponential superpositions are the interesting ones that can be created and manipulated by polynomial-in-$M$ complexity. Examples of exceptions are the oDJ problem [5] and phonebooks, where the database has exponential-in-$M$ complexity ($N=2^M$) but the special-purpose manipulations of the database require only polynomial-in-$M$ complexity. See more in Sections 3 and 4.

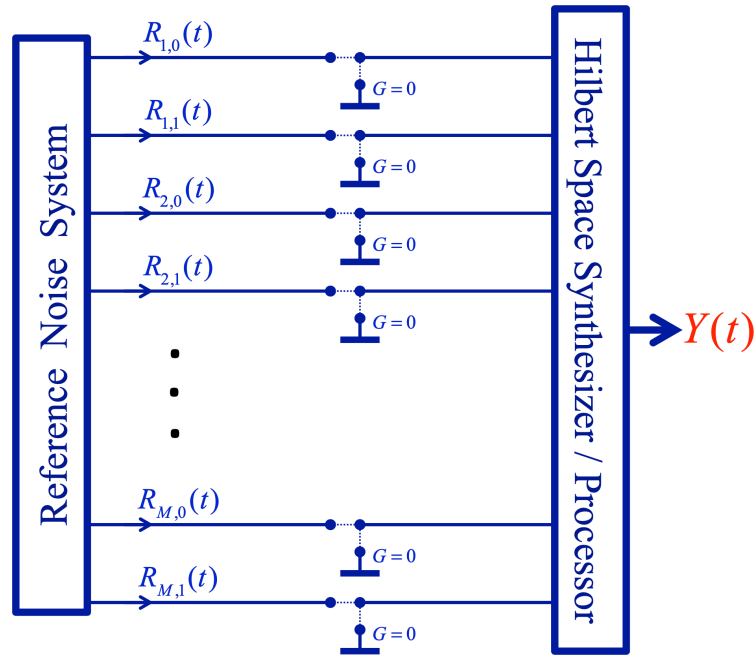

**Figure 5.** For string search (phonebook, etc.) some of the $R(t)$ amplitudes of the RNS system are forced to zero. In the circuit illustration of the logic structure in Figure 3, these represent "grounding" operations of those reference wires. Here, the INBL system circuitry has two-stage switches at each reference wire to represent the actuators for normal operation versus grounding. Such circuit illustrations make the assessment of the required complexity of the operations easy. In INBL systems running on a digital computer the algorithm contains the necessary operations, including the operations on the RNS signals, including forcing the zeros (grounding).

Some of the exponentially large, $O(N=2^M)$ superpositions can be treated by polynomial-in-$M$ complexity in INBL, which is also an essential feature of quantum computing. For example, the Universe $U(t)$, which is the superposition of *all* the binary numbers in the $M$ noise-bit system, can be created by adding the signals of the Low (0) and the High (1) bit values of each noise-bit and then multiplying these sums (Achilles heel operation [8,13-21]):

$$U(t) = \left[R_{10}(t) + R_{11}(t)\right]\left[R_{20}(t) + R_{21}(t)\right] \ldots \left[R_{M0}(t) + R_{M1}(t)\right] \tag{5}$$

Even though the time function $U(t)$ of the Universe can be set up by $M$ additions and $M-1$ multiplications, that is, by $2M-1$ elementary algebraic operations; when Equation (5) is expanded, it forms an exponentially large superposition, the sum of all the $2^M$ different product-strings (representing all the binary numbers) that this $M$ noise-bit system can form. For example, for $M=3$, it is:

$$\begin{aligned}U_{M=3}(t) = &R_{10}(t)R_{20}(t)R_{30}(t) + R_{11}(t)R_{20}(t)R_{30}(t) + R_{10}(t)R_{21}(t)R_{30}(t) + R_{11}(t)R_{21}(t)R_{30}(t) + R_{10}(t)R_{20}(t)R_{31}(t) + \\ &+ R_{11}(t)R_{20}(t)R_{31}(t) + R_{10}(t)R_{21}(t)R_{31}(t) + R_{11}(t)R_{21}(t)R_{31}(t)\end{aligned} \tag{6}$$

Other polynomial examples of representing special exponential superpositions by operations with polynomial complexity are:

The superposition of all the odd numbers and all the even numbers [21]:

$$U_{odd}(t) = R_{10}(t)\left[R_{20}(t) + R_{21}(t)\right] \ldots \left[R_{M0}(t) + R_{M1}(t)\right] \text{ and} \tag{7}$$

$$U_{even}(t) = R_{11}(t)\left[R_{20}(t) + R_{21}(t)\right] \ldots \left[R_{M0}(t) + R_{M1}(t)\right], \tag{8}$$

respectively.

Choosing the proper RNS is crucial [14,15] for the given application, for example, to avoid exponentially large bandwidth, or too many time-points with zero values of the superposition amplitudes, etc., see below.

*2.3. Complexity issues of the INBL hardware, and the exponentially large amplitudes of exponential superpositions*

A typical INBL framework is a binary computer (Turing machine) expanded with a true random number generator (TRNG). Note, that the universe, see Equation (5), and its fractions shown in Equations (7) and (8) represent $O[N=2^M]$, exponential-in-$M$, scale of possible amplitudes. However this is not a problem at all in a binary computer because these values can easily be carried by a CPU with polynomial-in-$M$ complexity, $O(M)$ bit resolution, see the example in Section-3.

On the other hand, as we have already mentioned, when the issue is representing $N=2^M$ data bits of *arbitrary* value in a database, the exponential $O[N=2^M]$ complexity of the *database generation* cannot be avoided in quantum computers, INBL or classical Turing machines.

Similarly, the core issue of the oDJ algorithm is as follows [5]. They start from a database of size $N=2^M$ (that is assumed to be doable with their unspecified hardware) and they show that they are able to solve the oDJ problem, (2.a)-(2.c), by their algorithm with $O[\log(N)=M]$ complexity. In this paper, we follow the same path.

# 3. Method: Implementation of the original Deutsch-Jozsa algorithm in noise-based logic

The oDJ paper [5] is focusing on the algorithmic solution of the problem and omits the functional hardware issues of how to create the database and how to output bit values at chosen addresses. In the present paper first these questions are addressed in the INBL framework before the algorithm that solves the oDJ problem is shown.

*3.1. The chosen Reference Noise System*

The RNS that we are using for the oDJ problem is part of the asymmetric, INBL scheme proposed in [14]. This is the same RNS as the one used for some of the earlier database related problems, namely the "Drawing from hats" [20] and the "phonebook search" problems [21,22]. It has the advantage for the oDJ case that the amplitude of its Universe is never zero. Its disadvantage is that it has a larger amplitude range than some other RNS schemes. This problem is manageable (including its Universe) by less than $4M$ bit resolution in a digital computer where it requires polynomial complexity with $4M$ bit resolution [14,20,21]. The reference noises have a periodic clock and they are two-state (binary, dichotomous) noises, called random telegraph waves (RTW), with ±1 amplitudes for the High (1) bit values and ±0.5 for the Low (0) bit values (with various binary amplitudes), see Figures 6,7 [20]. As the reference signals are independent RTWs, the products strings will also be RTWs with zero cross-correlation between different strings. The schemes have the advantage that binary multiplications and additions can be done accurately and economically by a digital computer while the bandwidth will remain the same during the multiplications in the product string.

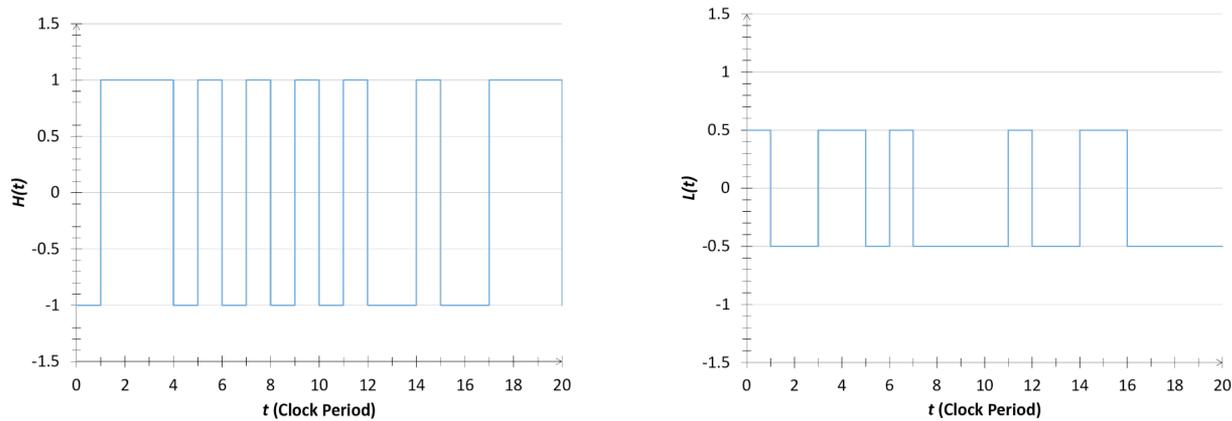

**Figure 6.** Computer simulations [20] of random telegraph waves carrying the High bit value (left) and the Low bit value (right) of a chosen noise-bit in the asymmetric INBL scheme. The clock is periodic. At the beginning of each clock period, an unbiased, truly random coin generates the choice between the two possible amplitude levels.

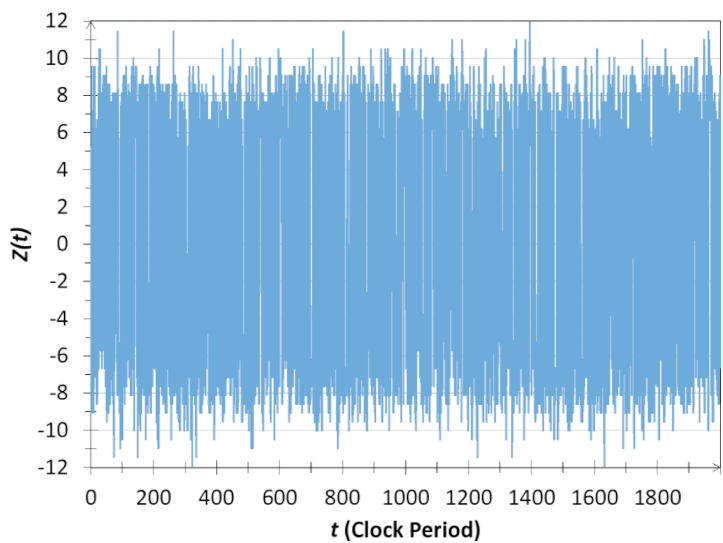

**Figure 7.** Computer simulation [20] of the $N = 32$ noise-bit universe in the asymmetric INBL scheme over 2000 clock periods. The time function $Y(t)$ of the universe, which represents the superposition of all integer numbers from 0 to 4294967295, is plotted in a logarithmically distorted way, defined by $Z(t) = \text{sign}\left[U(t)\right] \log\left[2^{32} |U(t)|\right]$, for better visibility of small and large variations.

*3.2. Implementation of the database by instantaneous noise-based logic*

Alice, who implements the INBL database, proceeds as follow. The carriers of the $N$ bits of the database are the $N$ $(=2^M)$ orthogonal product strings of the full universe generated by $M$ noise-bits.

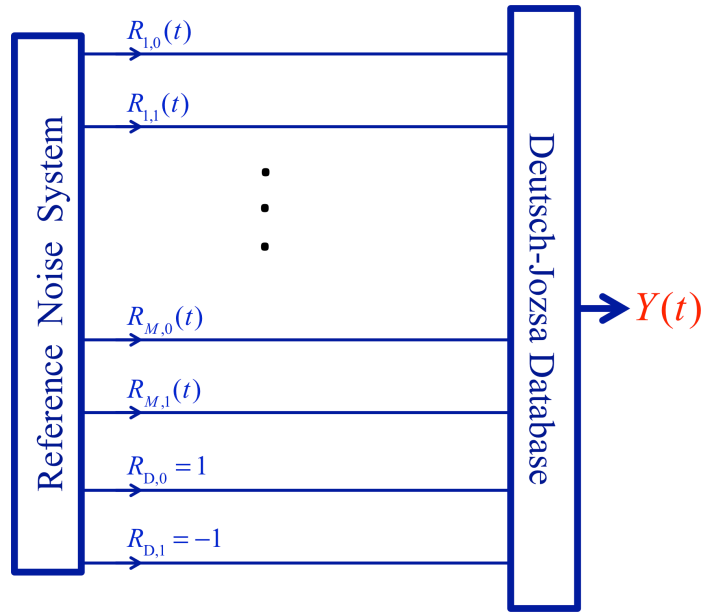

**Figure 8.** Circuit illustration of the logic scheme of the database. The *N* product strings (address strings) of the superposition (representing the universe of the *M* noise bits) carry the *N* carried/stored bit values of the database. Mathematically, the actual bit value is a positive or negative sign in front of the corresponding address strings. The expansion of the RNS by the $R_D$ wires carrying the +1 and -1 steady values represents the source for the carried/stored bit values used in the synthesizer (logic) circuitry of the database.

There is no need to create superpositions of the *carried* (stored) bit values. Thus they can be classical bits and only the carrier system requires INBL.

Thus in the INBL-based oDJ scheme the carried/stored database bit values are not represented by noise-bits but by constant values that are the $(M+1)$-th elements of the product strings. These constant signal components are +1 for logic Low (that is, binary bit value 0) and -1 for logic High (that is, binary bit value 1). For example, a database with $N=8$ classical bits requires an $M=3$ noise-bit system to carry these data. The output superposition of the database with the stored/carried bit values 1,0,1,1,0,0,1,1 are given as (compare with Equation (6)):

$$Y(t) = -R_{10}(t)R_{20}(t)R_{30}(t) + R_{11}(t)R_{20}(t)R_{30}(t) - R_{10}(t)R_{21}(t)R_{30}(t) - R_{11}(t)R_{21}(t)R_{30}(t) - R_{10}(t)R_{20}(t)R_{31}(t) - \\ +R_{11}(t)R_{20}(t)R_{31}(t) - R_{10}(t)R_{21}(t)R_{31}(t) - R_{11}(t)R_{21}(t)R_{31}(t) \quad , (9)$$

where, the negative elements are representing 1 (High) binary bit values. In Equation (9) we suppose that the order of the carried bit values is the same as the order of the product strings in the universe in Equation 6.

*3.2. Reading out data from the INBL database*

Bob, the user of the database must be able to output the bit value from any of the addresses. To output the stored bit value at a given memory address, an expanded version of the phonebook algorithm [21] is used. The protocol is as follows:

(i) "*Ground*" (force zeros to) the RNS at the inverse of the address string. Then only the address product string multiplied by the carried/stored bit value remains in the output superposition *Y*(t).

(ii) Multiply this output, *Y*(t), by $2^{2k}$, where *k* is the number of bits with Low values in the address string, and multiply also with the signal of the address string in order to rectify the address string signal, as the square of any product string is a positive constant. Therefore, these steps eliminate the time dependence, and scale up the absolute value of the result to 1. The result will be the bit value of the carried/stored bit at this address, that is, either +1 (Low) or -1 (High).

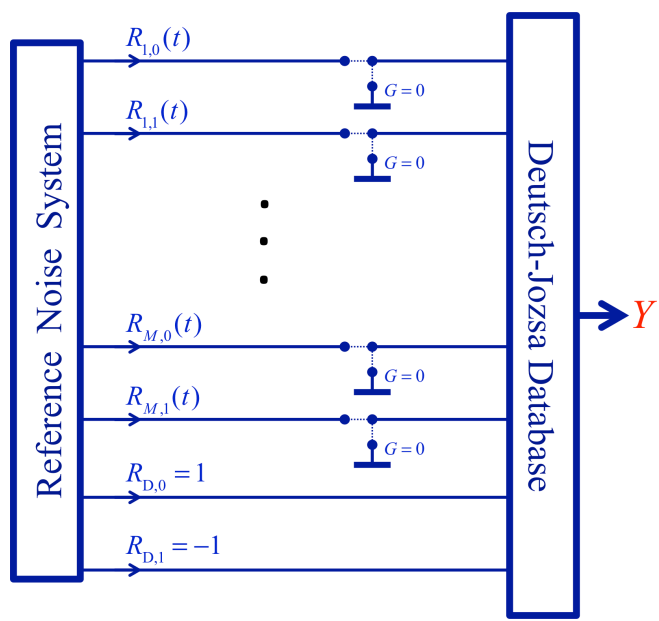

**Figure 9**. Circuit illustration of the logic operation of the grounding of the RNS signals. The inverse of the binary address of the memory cell to be outputted must be grounded at the reference wires. In other words, the inverse of the address must be forced to zero in the RNS. This operation is illustrated by the 2-stage switches in the figure. The resulting output *Y* contains only the string that carrying the bit value in question. This output must be multiplied by the signal of the address string to rectify the address noise, and also by $2^{2k}$, where *k* is the number of bits with Low values in the address string, in order to project the relevant +1 or -1 value of the memory bit in question.

For example, to output the carried/stored bit value at the address 010 in the database in Equation 9, proceed as follows:

(a) Ground the RNS elements 101, that is, $R_{1,1}(t) = 0$, $R_{2,0}(t) = 0$, $R_{3,1}(t) = 0$. Then the output becomes

$$Y_{010}(t) = -R_{10}(t)R_{21}(t)R_{30}(t), \qquad (10)$$

where

$$|Y_{010}(t)| = \frac{1}{4}, \qquad (11)$$

because the address string contains two bits values with zero, see Figure 6.

(b) Then multiply $Y_{010}(t)$ by $16 R_{10}(t)R_{21}(t)R_{30}(t)$. The result is the carried/stored bit value -1 (H) at the address 010.

$$16 R_{10}(t)R_{21}(t)R_{30}(t)Y_{010}(t) = -16 R_{10}(t)R_{21}(t)R_{30}(t)R_{10}(t)R_{21}(t)R_{30}(t) = -1. \qquad (12)$$

Finally, a remark that is not the topic of the present paper but it is about potential applications: this system is able to perform more than a typical database can do because certain parallel logic operations can be performed on the whole system (for example, NOT, CNOT, etc., gate operations simultaneously on all the $2^M$ product strings [15,19-21]). However, such operations must be done with precautions, see [22].

### 3.3. Running the original Deutsch-Jozsa problem on the exponential database

To run the oDJ problem and to make a deterministic decision about it, Bob needs to execute an $\mathrm{O}[\log(N)]$ number ($2M$) of single switching operations and a single amplitude measurement (read-out) at the output:

-Bob switches (forces) all the $2M$ different RNS signals to the same constant value of 1, see Figure 10.

-Bob reads (or measures) the output amplitude $Y$.

Then, obviously:

(A) If the data are all logic Low (bit value 0) then $Y = 2^M$.

(B) If the data are all logic High (bit value 1) then $Y = -2^M$.

(C) If the data are 50-50% High and Low then $Y = 0$.

Note: as we have already mentioned, the natural hardware for INBL is a digital computer, where the switching of the $2M$ different RNS signals of the carrier to the same the constant value of 1 and the reading out of the $|Y| = 2^M$ value require only $\mathrm{O}(M) = \mathrm{O}[\log(N)]$ complexity for Bob: $2M$ single switching operations, and reading the output with $M+1$ classical bit resolution.

Concerning the oDJ statements, see Section 2.1, the solution by Bob is as follows:

(D) Whenever $|Y| \neq 2^M$, the statement (2.a) is true;

(E) Whenever $Y \neq 0$, the statement (2.b) is true;

(F) Whenever $|Y| \neq 2^M$ AND $Y \neq 0$, statement (2.c) is relevant. Note, as we have mentioned in Section 2.1, this situation is completely excluded in the simplified Deutsch-Jozsa problem due to its (2.d) condition.

In conclusion, the original Deutsch-Jozsa problem is deterministically decided by Bob's operations with $\mathrm{O}[\log(N)]$ complexity.

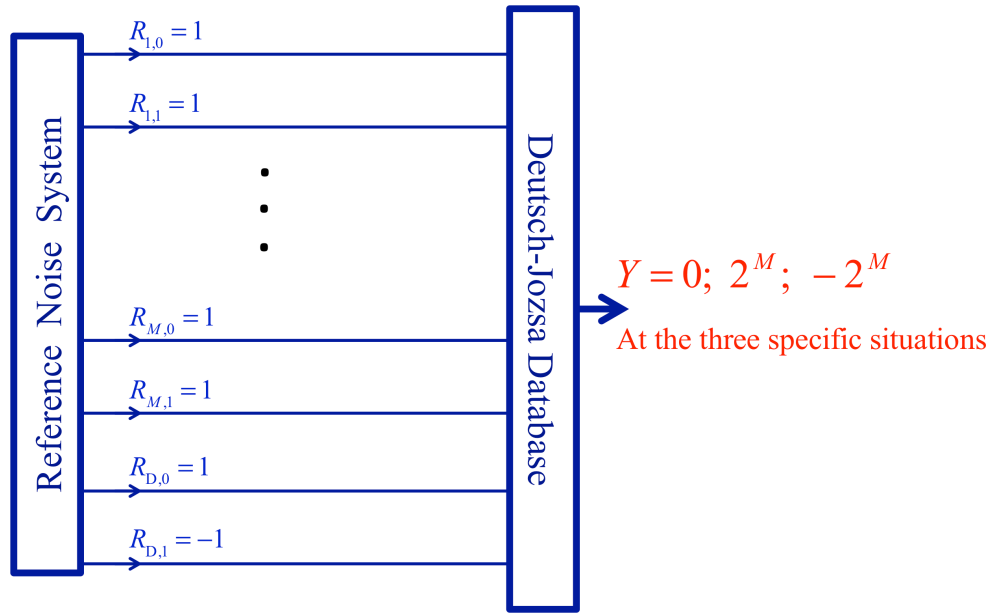

**Figure 10.** Bob solves the original Deutsch-Jozsa problem by operations of O[log(*N*)] complexity. For details, see the text.

# 4. Discussion

*4.1 Turing machine with a random coin: quantum or classical?*

A natural question emerges regarding quantum supremacy: Is a random coin a quantum or a classical device? True random number generators that utilize the Born rule of quantum theory are available on the market. However, electronic white noises, such as thermal noise and shot noise are also excellent sources of true randomness. On the other hand, these noise processes also have quantum physics at their microscopic origin. An example is the phonon scattering of the electrons. See more in [18].

Yet, the further discussions of this intriguing question can be avoided in this paper due to the findings described in section 4.2

*4.2 Implementing the original Deutsch-Jozsa problem without a random coin (noise)*

Unexpectedly, the scheme in Figure 10 implies that there is yet another realization, even without using noises in the RNS. Bob can introduce the addressing switches in Figure 9 into the circuitry in Figure 10 and, with the same hardware framework, he is able to build the same database and perform both the readout (II) and oDJ decision (III) operations even when the noise sources (reference signals) driving the system are replaced by steady +1 values, see Figure 11.

When the database is idle, that is, when no readout is happening, the output is readily showing Bob the solution of the oDJ problem in the same way as it is described in Section 3.3.

To read out the bit value of a chosen address, Bob, simply follows the same grounding protocol (forcing zero values to the inverse of the address) as described in Section 3.2. This time, there is no need to multiply the *Y* output with anything because all the reference values are now +1, except the reference wire $R_{D,1}$ for the carried bit values of -1, see Figure 11. The output *Y* will be +1 or -1 depending on the value of the stored bit at the address. For example, when outputting the stored bit value at the 010 address as it is shown above, after the grounding operation of the 101 RNS values, Equation 10 becomes:

$$Y_{010} = -R_{10}R_{21}R_{30} = -1 \qquad (13)$$

because $R_{10} = R_{21} = R_{30} = 1$.

The only lost functionality with this simplified system is the parallel computational abilities over the database [15,19-22], which is however not part of the original Deutsch-Jozsa problem.

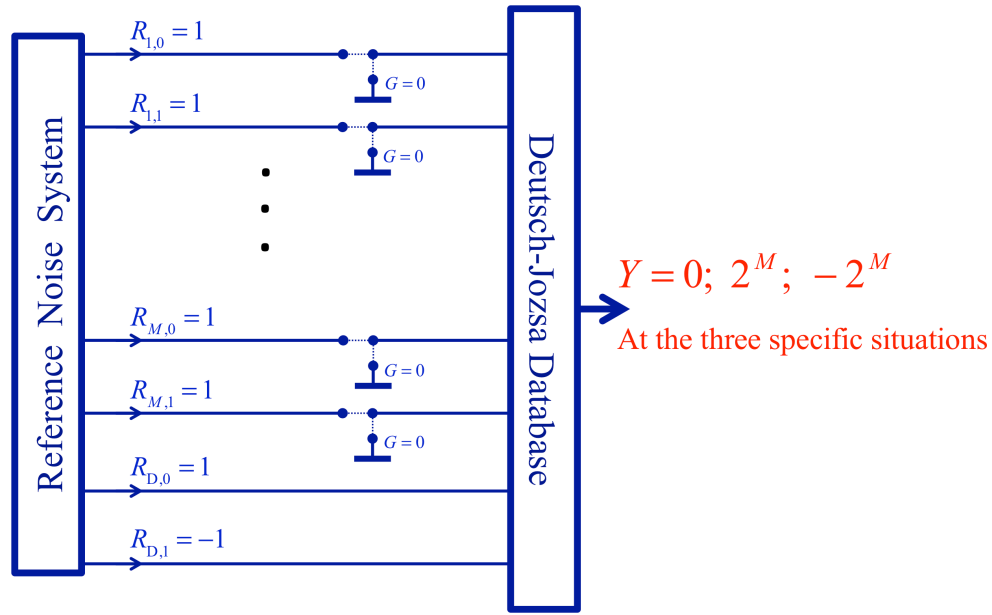

**Figure 11.** A classical physical database and processor with the Deutsch-Jozsa problem implemented. Bob can output the bit value at any location by $M$ switchings (grounding the $M$ reference wires representing the inverse of the address), that is, by O[(log$N$)] polynomial complexity. When the database is idle, that is, when no single bit values are outputted, the output directly shows the solution of the oDJ problem, see Section 3.3. Reading the solution requires $M+1$ bit resolution, which means O[(log$N$)] complexity.

# 5. Conclusion

It is observed that, in the *original* Deutsch-Jozsa problem, for arbitrary data of size $N$, Alice (who sets up the database) has to have O($N$) hardware/time complexity, *in both quantum and classical computers*.

(I) Alice constructs and sets up a database with $N=2^M$ bit capacity where the information carrier and the addressing system are represented by the Universe of an instantaneous noise-based logic framework of $M = \log_2(N)$ noise-bits, and the output is the superposition of the whole system (carriers+data), see Section 3.1.

(II) Bob can project any single bit of this database to the output and, to do that, he needs only polynomial-in-$M$, O[log($N$)], hardware/time complexity of the addressing. That means the database readout is effective, see Section 3.2.

(III) In this system, Bob can deterministically solve the oDJ problem with polynomial-in-$M$, O[log($N$)], hardware/time complexity. That means a classical Turing computer with an additional truly random coin and a special-purpose algorithm, is able to solve the oDJ problem with the same complexity as the solution by the oDJ algorithm [5] in a quantum computer, see Section 3.3.

(IV) Then, see Section 4, it is realized that Bob, using the same hardware framework, is able to perform both the (II) and (III) operations even when the noise sources (reference signals) driving the addressing part of the system are replaced by steady +1 values. The only lost functionality is the computational abilities within the database, which is however not part of the oDJ problem.

(V) In conclusion, the existence of a deterministic solution by Bob with O[log($N$)] complexity in a classical physical system is an indication that the oDJ problem and its quantum algorithm cannot prove quantum supremacy.

Yet, the original Deutsch-Jozsa problem remains an interesting historical marker of the development of quantum computer theory.

Finally, we note that that the solution of the simplified DJ problem by NBL is a challenging open question. Regarding the setting up of the oracle/database for the simplified DJ problem: as it is shown by equations 5,7,8, certain exponential superpositions can be set up by polynomial complexity, which is encouraging. However, the explorations of the solution of the simplified DJ problem remain a future goal.

**Acknowledgments**
The author is grateful to Tamas Horvath (Univ. Bonn, Fraunhofer IAIS) for informing him about the Deutsch-Jozsa problem and a discussion about oracles. The author is also grateful for Horace Yuen (Northwestern Univ., Chicago) for discussions about the quantum informatics aspects of the simplified Deutsch-Jozsa problem.

**Ethical Statement**
Research on humans must include a statement detailing ethical approval and informed consent. Research using animals must adhere to local guidelines and state that appropriate ethical approval and licenses were obtained.
N/A

**Funding Statement**
N/A

**Data Accessibility**
N/A

**Competing Interests**
I have no competing interests.

**Authors' Contributions**
N/A